\documentclass[12pt]{article}
\usepackage{latexsym}
\usepackage{graphicx}

\thicklines                         
\long \def \blockcomment #1\endcomment{}

\begin{document}           

\baselineskip=0.33333in
\begin{quote} \raggedleft  CHAR-02
\end{quote}

\large
\baselineskip=0.33333in

\begin{center}{\bf The Crucial Role of Inert Source \\
in the Magnetic Aharonov-Bohm Effect}
\end{center}
\begin{center}E. Comay$^*$
\end{center}

\begin{center}

Charactell Ltd. \\
P.O. Box 39019, Tel Aviv 61390, Israel
\end{center}

\vglue 0.5in
\noindent
\vglue 0.2in
\noindent
Abstract:

The role of the inert magnetic source used in the Tonomura experiment
that has confirmed the magnetic Aharonov-Bohm effect is discussed. For
this purpose, an analysis of a thought experiment is carried out.
Here the permanent magnet is replaced by a classical
source which is made of an ideal coil. A detailed calculation
of this noninert source proves
that in this case the effect disappears. This outcome provides
another support for the crucial role of an inert source in the
Aharonov-Bohm effect. A new aspect of quantum nonlocality is
pointed out.

\newpage


\noindent
{\bf 1. Introduction}
\vglue 0.33333in

As is well known, the Lorentz force (see [1], p. 51) is the
classical equation of motion of a charged particle. This equation
is written in terms of electromagnetic fields. Here the potentials
are auxiliary mathematical quantities that may (or may not)
be used for solving problems.
On the other hand, fundamental
quantum mechanical equations (like the Schroedinger
and the Dirac equations) depend explicitly on
electromagnetic potentials. Differences between classical and
quantum mechanical equations may arise from this structure of the theories.
However, the
consistency of these two kinds of theories is proved by Ehrenfest
theorem (see [2], pp. 25-27, 137, 138)
which shows that the classical limit of
quantum mechanics is consistent with classical physics. This
theorem settles the main problem and shows that classical
physics and quantum mechanics can live side by side.
Thus, classical physics holds only for experiments that belong to a
the classical limit of quantum mechanics
whereas quantum mechanics holds for a much larger
set of experiments. The main advantage of classical physics is
that its equations are much simpler than the corresponding equations
of quantum mechanics.

In their work [3],
Aharonov and Bohm (AB) examine phase properties of a quantum mechanical
charged particle that moves in field free region where the external
potentials do not vanish. Evidently, vanishing fields guarantee
that a classical experiment carried out under these conditions should yield
null results. On the other hand, AB
argue that the dependence of the phase on the potentials should yield a
phase shift in cases where a quantum mechanical charged particle moves in a
multiply connected field free region. This phase shift affects the
interference pattern of such a quantum mechanical charge.

It is well known that microscopic phenomena generally cannot be explained
by classical physics. Some macroscopic phenomena, like superconductivity,
superfluidity and EPR related experiments
are also outside the scope of classical physics.
Similarly, an AB experiment that measures phase difference of a
massive particle, has no classical analog. In this sense, it provides
another kind of a macroscopic quantum mechanical effect because the
interference pattern depends on fields that are quite far from all
possible trajectories of the electron.

The AB ideas about the magnetic AB effect have been confirmed by the work
of Tonomura et al.\ who have constructed an appropriate experiment [4].
(Below, the experiment described in [4] is called the Tonomura experiment.)
This experiment uses a ring of a single domain of a ferromagnetic
material. Evidently, the magnetic ring behaves like an inert object
throughout the experiment. A general remark on the importance of this
property of the magnetic component of Tonomura's experiment
has been pointed out in the literature [5,6].
The main purpose of the present work is to analyze an experiment
where the Tonomura quantized magnet is replace by an analogous
classical coil and thereby, to show that the existence of an inert
magnetic source is crucial for a nonvanishing AB phase shift. The
result also shows a new aspect of quantum mechanical nonlocality.

Expressions are written in units where $\hbar = c = 1$.
The second section describes the Tonomura experiment. Calculations of
a classical analog of the Tonomura experiment are presented in the
third section. Concluding remark on consequences of the analysis
are included in the last section.

\vglue 0.66666in
\noindent
{\bf 2. The Tonomura Experiment}
\vglue 0.33333in

\begin{figure}[b]
  \centering
    \includegraphics[width=3.5cm,height=4.8cm]{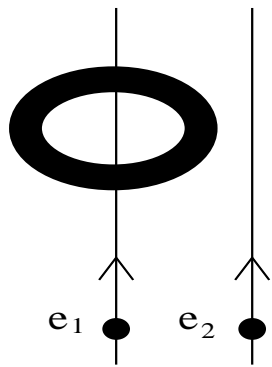}
  \caption{\em The main elements of the Tonomura experiment (see text). }
\end{figure}

The Tonomura device aims to test the phase shift predicted by
the magnetic AB effect,
where the electrons move in an external nonsimply connected
field free region. A phase dependent
interference pattern of an electronic beam is obtain by means of
electron holography.
The beam passes near a circular ring made of a ferromagnetic material
(see fig. 1).
The ring is a (nearly) perfect magnetic single domain. It is
covered by a superconducting material and by copper. This arrangement
prevents the beam's electrons from entering the region where the
magnetic field does not vanish. The usefulness of dividing the beam
into two subbeams is explained later. One subbeam $G_1$ passes
through the ring's inner region and the second subbeam $G_2$ passes
at the ring's outer region (see fig. 1). Both subbeams move
in a magnetic field free region.

For phase-difference calculation, one has to examine
the development of the action along possible
trajectories of the subbeams.
Below, the electron whose interference is analyzed is called
the traveling electron. Quantities pertaining to the traveling electron
are denoted by the subscript $(e)$. Other quantities pertain to the
magnet.

Let us calculate the rate of phase accumulated.
Thus, the action is the time integral of the Lagrangian of this system
\begin{equation}
{\mathcal L}_{total} = {\mathcal L} + {\mathcal L}_{(e)} -
e{\mathbf v}_{(e)} \mathbf {\cdot A}.
\label{eq:LAG}
\end{equation}
The terms on the right hand side
represent the Lagrangian of the magnet, of the traveling electron
and of their interaction, respectively. (Here $-e$ denotes the
electronic charge.) Obviously, the state of the ferromagnetic ring is
independent of the traveling electron. Therefore,
the first term of $(\!\!~\ref{eq:LAG})$ makes the same contribution
to all possible trajectories pertaining to
the electronic beam. The same is true for
the second term of $(\!\!~\ref{eq:LAG})$, since, due to the field
free region where the electronic beam moves, the self (kinetic)
energy of the electron is constant.

In order to calculate the required
interference pattern, one must have an
expression for {\em phase difference} accumulated on
any pair of possible trajectories of the electronic beam. Now,
due to the constant value of the first and
the second terms of $(\!\!~\ref{eq:LAG})$, these terms make no
contribution to the phase difference.
Let $l_1$ and $l_2$ denote two trajectories that
begin at the beam's origin and meet at a point on the screen where
the interference is measured. Integrating $(\!\!~\ref{eq:LAG})$
on time, writing ${\mathbf v}\,dt = d{\mathbf x}$
and using vector analysis, one takes the
last term of $(\!\!~\ref{eq:LAG})$
and obtains the required phase difference
\begin{eqnarray}
\Delta \Phi & = &
-\int _{l_1} e\mathbf {A \cdot} d \mathbf x +
 \int _{l_2} e\mathbf {A \cdot} d \mathbf x
\nonumber \\
         & = &
\oint _l  e\mathbf {A \cdot} d \mathbf x
\nonumber \\
         & = &
\int _s  e(\nabla \times \mathbf {A}) \mathbf \cdot d \mathbf s
\nonumber \\
         & = &
\int _s  e \mathbf {B \cdot }d \mathbf s
\label{eq:DELTAPHI}
\end{eqnarray}

This result means that the phase shift of two possible trajectories of
the traveling electron depends on the magnetic flux passing
through an area whose boundary is determined by the closed line defined by
$l_1$ and $l_2$. Here a usage of the two sets of beams $G_1$ and $G_2$
yields straightforwardly the required result. Thus, if both
$l_1$ and $l_2$ belong to the same set then no magnetic flux is found
and a null phase shift is obtained.
On the other hand, the {\em same} nonvanishing
phase shift is obtained for two trajectories that belong to
different sets.

The Tonomura experiment [4] has confirmed the AB's prediction [3] which is
described in the previous lines.

\vglue 0.66666in
\noindent
{\bf 3. A Classical Analog of the Tonomura Experiment}
\vglue 0.33333in

The following discussion proves the crucial role of an
inert source in a test of
the magnetic AB effect [3].
Let us consider a thought experiment where the
Tonomura's magnetic ring is
replaced by a coil having these properties. The coil is a
closed pipe which is made of an insulating material (see fig. 2).
The pipe contains a uniformly charged incompressible liquid that flows
frictionlessly along the pipe. The pipe itself is covered uniformly
with an appropriate density of electric charge of the opposite sign.
Thus, outside
the pipe there is no electric field and a ring of a magnetic
flux exists at the coil's inner part. This is a
"classical analog" of Tonomura magnet.

\begin{figure}[t]
  \centering
    \includegraphics[width=5.0cm,height=4.8cm]{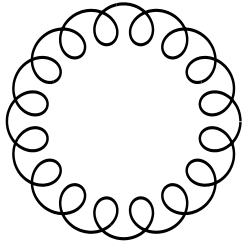}
  \caption{\em A closed circular coil (see text). }
\end{figure}

Let us compare the
interference pattern of Tonomura experiment with that
which is expected to be found in an experiment with the classical source
described herein.
Like in the standard presentations of the AB effect [3],
the following calculations
are carried out within the nonrelativistic limit.
The calculations are analogous to those of [5,6] and
the result provides a
further justification for the indispensable role of an inert source
in the AB effect.

Let $a$ denote the inner radius of the pipe and $R$ the radius of the
inner part of the coil where the magnetic field does not vanish.
The relation $a\ll R$ simplifies the calculations presented below.
The symbols $\rho $ and $v$ denote the linear charge density and
the velocity
of the charged liquid flowing along the pipe, respectively.

Let us examine the Lagrangian $(\!\!~\ref{eq:LAG})$
for the case where the
coil replaces the permanent magnet.
The calculations take a
simpler form if the coil is regarded as a dense assembly of
identical loops, each
of which contains the same uniformly
charged liquid that flows at the same
velocity $v$. Thus, the problem of
the traveling electron and one loop is analyzed (see
fig. 3). The result for the entire coil will be derived
from this analysis.

\begin{figure}[t]
  \centering
    \includegraphics[width=2.5cm]{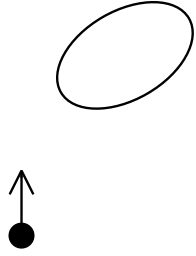}
  \caption{\em The traveling electron and one loop (see text). }
\end{figure}

The loop's vector potential is obtained from an integration on the
charged liquid flowing along the loop
\begin{equation}
{\mathbf  A} = \oint \frac {\rho v}{r}d\mathbf l,
\label{eq:ALOOP}
\end{equation}
where $r$ denotes the distance from the line element $d\mathbf l$ to
the field point where $\mathbf A$ is calculated.
Thus, the interaction term of $(\!\!~\ref{eq:LAG})$ is cast into the
following form
\begin{equation}
-e{\mathbf v}_{(e)} \mathbf {\cdot A} =
-e\rho v \oint \frac {\mathbf v_{(e)}\mathbf \cdot d \mathbf l}{r}
\label{eq:ANEW}
\end{equation}

Now, due to the insulating material of the pipe, the charge that covers
it is static throughout the experiment. Therefore, its self energy is
a constant of the motion and  {\em it also does not
screen the fields of the traveling electron.} The last point means that
the kinetic energy $T$ of the rotating liquid
as well as the associated Lagrangian
may change during the process [7]. Furthermore, for the
rotating liquid the ratio of the charge density to the mass density
is very very small (relative to the corresponding ratio of an electron).
Hence, the calculation is simplified if the
negligible change in the liquid's velocity is ignored. For a
time instant $t$, one uses vector analysis and Maxwell equations and
finds the change of
the kinetic energy of the rotating charged liquid
\newpage
\begin{eqnarray}
\Delta T & = &
\int _{-\infty }^t \rho v[\oint \mathbf E _{(e)}\mathbf \cdot d\mathbf l]dt
\nonumber \\
         & = &
\int _{-\infty }^t \rho v[\int_S (\nabla \times \mathbf E _{(e)})
\mathbf \cdot d\mathbf s]dt
\nonumber \\
         & = &
-\int _{-\infty }^t \rho v[\int _S \frac {\partial \mathbf B _{(e)}}
{\partial t} \mathbf \cdot d\mathbf s]dt
\nonumber \\
         & = &
-\rho v\int _S \mathbf B_{(e)} \mathbf \cdot d\mathbf s
\nonumber \\
         & = &
-\rho v\int _S \nabla \times \mathbf A_{(e)} \mathbf \cdot d\mathbf s
\nonumber \\
         & = &
-\rho v\oint  \mathbf A_{(e)} \mathbf \cdot d\mathbf l
\nonumber \\
         & = &
e\rho v \oint \frac {\mathbf v_{(e)}\mathbf \cdot d \mathbf l}{r}
\label{eq:DEK}
\end{eqnarray}
This calculation proves that
$(\!\!~\ref{eq:ANEW})$ and $(\!\!~\ref{eq:DEK})$ cancel each other.
Thus, no phase shift is found for one loop of current. It follows
that the combined interaction of the traveling electron with the
coil and its field makes no contribution to the phase shift.

The present experiment has the same
magnetic flux and the same multiply connected
field-free region as that of the Tonomura experiment. However,
as stated above,
an examination of $(\!\!~\ref{eq:ANEW})$ and $(\!\!~\ref{eq:DEK})$
proves that their contribution to the rate of phase accumulated
cancel each other. It follows that, unlike the inert single domain
used in the Tonomura experiment,
a classical magnetic source does not alter the interference pattern.

\vglue 0.66666in
\noindent
{\bf 4. Concluding Remarks}
\vglue 0.33333in

Several effects proving the macroscopic scale of
quantum mechanics and of its
nonlocality are mentioned in the introduction.
The magnetic AB effect, whose existence is demonstrated by the
Tonomura experiment certainly belongs to this category. Indeed, scaling
length by atomic size, one finds that the distance between
the electron's path and the magnetic field is very large.
In spite of this fact, the electronic state is
affected by the relatively remote
magnetic field and "remembers" it even for the macroscopic
distance between the interference region and the magnetic source.

The analysis presented above shows a new aspect of quantum mechanical
nonlocality. Thus, the electronic interference depends not only on
the magnetic field as is, but also {\em on the specific device that
produces this magnetic field}. In the case of the ferromagnetic single
domain used in the Tonomura experiment, the magnetic AB
effect exists. On the
other hand, if the {\em same} magnetic field is produced by
the ideal coil described above,
then the phase shift disappears and the (macroscopically far)
interference pattern changes. Now, the traveling electron touches
neither the magnetic field nor the device that produces this field. However,
the interference pattern proves that
the electronic beam interacts
not only with the magnetic field which it does not touch but also
with the device that produces this field.

The inherent nonlocality of the AB effect is summarized in the
following statements. The interference pattern is an assembly of
dots, each of which is created by the collision of one
electron with the screen. The structure of the interference
pattern reflects the probability of finding the traveling
electron at any small area on the screen.  This probability is the
absolute value of the square of the overall amplitude and this
amplitude is obtained by taking the appropriate sum
of the contribution of all
relevant trajectories of the traveling electron. This sum is
very sensitive to the phase factor and it determines the
location of constructive and destructive interference regions. The phase
accumulated on any possible trajectory is the action $(\hbar = 1)$
obtained as the time integral of the Lagrangian
$(\!\!~\ref{eq:LAG})$. Now, the Lagrangian $(\!\!~\ref{eq:LAG})$
depends on {\em all} coordinates of the system. Thus, the
source contributes to the phase accumulated on every trajectory
and the traveling  electron "remembers" it. Now, a dot on the
screen is certainly a local property created by the collision
of the traveling electron with the screen. However, this local
property is affected by the source and its magnetic field
even if the traveling electron has
never made any contact with them.

Other conclusions can also be inferred from the discussion presented above:
\begin{itemize}
\item[{1.}] The existence of a phase shift crucially depends on a
source that behaves as an inert object throughout the experiment.
\item[{2.}] The multiply-connectedness topology is not sufficient
for having an AB effect. This requirement must be augmented by
demanding the usage of an inert source of the magnetic field.
\end{itemize}

\vglue 0.5in

References:

\begin{itemize}
\item[{$^*$}] Email: elicomay@post.tau.ac.il  \\
\hspace {1.5cm}
Internet site: http://www.tau.ac.il/$\sim $elicomay

\item[{[1]}] L. D. Landau and E. M. Lifshitz, {\em The Classical
Theory of Fields} (Elsevier, Amsterdam, 2005).
\item[{[2]}] L. I. Schiff, {\em Quantum Mechanics} (McGraw-Hill, New York,
1955).
\item[{[3]}] Y. Aharonov and D. Bohm, Phys. Rev. {\bf 115}, 485 (1959).
\item[{[4]}] A. Tonomura et al., Phys. Rev. Lett. {\bf 56}, 792 (1986).
\item[{[5]}] E. Comay, Phys. Lett. {\bf A250}, 12 (1998).
\item[{[6]}] E. Comay, Phys. Rev. {\bf A62}, 042102 (2000).
\item[{[7]}] The usefulness of the specific structure of the coil
is now clear. Indeed, here one only needs to carry out the straightforward
calculation $(\!\!~\ref{eq:DEK})$. On the other hand, in the case of a
metallic coil, the traveling electron induces changes of the coil's charge
density and current. Thus, for a metallic coil, one must cope
with the tedius task of calculating the time dependence of these
quantities and of the associated self-energy of the coil.

\end{itemize}

\end{document}